\documentclass[manuscript]{acmart}
\AtBeginDocument{%
  }

\copyrightyear{2025}
\acmYear{2025}
\begin{document}

\title{Intelligent Interaction Strategies for Context-Aware Cognitive Augmentation}
\begingroup
\renewcommand\thefootnote{}\footnote{
 This paper was presented at the 2025 ACM Workshop on Human-AI Interaction for Augmented Reasoning (AIREASONING-2025-01). This is the authors’ version for arXiv.}
\endgroup

\author{Xiangrong (Daniel) Zhu}
\email{xzhu744@connect.hkust-gz.edu.cn}
\orcid{0009-0001-2801-2199}
\author{Yuan Xu}
\email{yuanxu@hkust-gz.edu.cn}
\orcid{0009-0004-0811-9505}
\author{Tianjian Liu}
\email{tianjianl@hkust-gz.edu.cn}
\orcid{0009-0006-9482-9689}
\affiliation{%
  \institution{The Hong Kong University of Science and Technology (Guangzhou)}
  \city{Guangzhou}
  \country{China}
}

\author{Jingwei Sun}
\email{sunjw12@lenovo.com}
\orcid{0000-0002-9737-1056}

\author{Yu Zhang}
\email{zhangyu29@lenovo.com}
\orcid{0000-0001-7298-1694}
\affiliation{%
  \institution{Lenovo Research}
  \city{Beijing}
  \country{China}
}

\author{Xin Tong}
\authornote{corresponding author}
\affiliation{%
  \institution{The Hong Kong University of Science and Technology (Guangzhou) and The Hong Kong University of Science and Technology}
  \city{Guangzhou}
  \country{China}}
\email{xint@hkust-gz.edu.cn}

\renewcommand{\shortauthors}{Zhu et al.}

\begin{abstract}
Human cognition is constrained by processing limitations, leading to cognitive overload and inefficiencies in knowledge synthesis and decision-making. Large Language Models (LLMs) present an opportunity for cognitive augmentation, but their current reactive nature limits their real-world applicability. This position paper explores the potential of context-aware cognitive augmentation, where LLMs dynamically adapt to users' cognitive states and task environments to provide appropriate support. Through a think-aloud study in an exhibition setting, we examine how individuals interact with multi-modal information and identify key cognitive challenges in structuring, retrieving, and applying knowledge. Our findings highlight the need for AI-driven cognitive support systems that integrate real-time contextual awareness, personalized reasoning assistance, and socially adaptive interactions. We propose a framework for AI augmentation that seamlessly transitions between real-time cognitive support and post-experience knowledge organization, contributing to the design of more effective human-centered AI systems.
\end{abstract}





\maketitle

\section{Introduction}

Human cognition is constrained by fundamental processing limitations, creating a gap between perception and reasoning~\cite{goette2020information}. While sensory systems acquire information at approximately $10^9$ bits per second, cognitive processing operates at only about $10$ bits per second~\cite{zheng2025unbearable}. This bottleneck limits individuals' ability to synthesize, connect, and apply knowledge efficiently, often leading to cognitive overload and decision-making biases. LLMs offer a promising solution by augmenting human reasoning, but their current reactive nature restricts real-world applicability~\cite{zhou2024larger}. To bridge this gap, we propose \textbf{context-aware cognitive augmentation}, where LLMs dynamically adapt to users' cognitive states and task environments, proactively assisting in information processing, structured reasoning, and decision-making. By integrating multi-modal signals and real-time context, LLMs can move beyond passive knowledge retrieval to \textbf{actively supporting human cognition} in complex, information-rich settings.

Human cognition follows a structured process that filters, interprets, and applies information~\cite{hilbert2011world}. Perception channels sensory inputs (e.g., vision, audition, touch), but selective attention limits what enters cognitive processing. Cognitive mechanisms then organize and integrate the selected information, constrained by working memory capacity and processing bandwidth~\cite{lewis2020retrieval}. Finally, individuals apply processed knowledge in real-world tasks, yet memory retrieval often suffers from decay and reconstruction, affecting accuracy and efficiency~\cite{roberts2023knowledge}. Understanding these limitations is critical for designing AI-driven cognitive support systems that enhance human reasoning.

However, supporting human cognition effectively requires more than just providing access to vast amounts of information. The ability to filter, contextualize, and apply relevant knowledge in real time is crucial for mitigating cognitive constraints. Despite their potential, existing LLMs lack contextual awareness, providing one-size-fits-all responses rather than adapting to users' cognitive processes~\cite{baek2024knowledge}. Effective augmentation requires LLMs to be proactive, leveraging real-time contextual information to tailor responses and facilitate structured reasoning. 

Our preliminary study tends to investigate two research questions:
\begin{itemize}
    \item How do individuals sense, process, and apply information in real-world tasks, and what cognitive support do they expect?
    \item How should AI-driven cognitive support systems be designed to enhance knowledge augmentation?
\end{itemize}

To explore these questions, we conducted a think-aloud study in an exhibition setting. Participants navigated exhibits while capturing and synthesizing information for later use. This scenario provided a rich multi-modal environment to observe cognitive behaviors. Findings reveal challenges in structuring contextual information, retrieving knowledge, and leveraging digital tools for context-aware, structured, and proactive engagement. Based on these insights, we outline a \textbf{context-aware cognitive augmentation approach} that allows LLMs to adjust responses dynamically based on user interactions and cognitive needs. This work contributes by identifying key cognitive challenges in knowledge augmentation, offering empirical insights into users' expectations for AI-driven support, and suggesting design considerations for integrating context-aware reasoning into LLM-based systems.

\section{Related Work}

\subsection{Context-Aware in AI-Augmented System}
With the current information explosion~\cite{Kitsuregawa2010Special}, the amount of data we can access keeps increasing. A core objective of modern AI-augmented systems is to help people collect and process information within limited attention spans and working memory capacities~\cite{zulfikar2024memoro}. Chen et al. \cite{chen2023gap} proposed LangAware, which uses in-situ contextual data to filter unnecessary notifications, allowing people to reduce cognitive burdens in different life scenarios, thus achieving better human-machine collaboration.

Another approach involves user embeddings, where AI systems create numerical representations of a user's interaction history and cognitive style~\cite{ning2024user}. These embeddings allow LLMs to adjust response styles dynamically, improving personalization. However, they lack structured knowledge dependencies, such as how a user's beliefs evolve over time or how exposure to new perspectives influences reasoning~\cite{zhang2024personalization}.

Knowledge graphs provide a structured method for representing user information, encoding their knowledge, preferences, and prior interactions into a graph-based system. AI models can query these graphs to retrieve personalized insights, which has been particularly effective in educational AI systems~\cite{rasheed2024knowledgegraphs}. However, applying them to support knowledge-intensive tasks still poses an open research challenge~\cite{lewis2020retrieval}.

\subsection{LLM-Empowered Cognitive Augmentation}
Knowledge Augmentation plays a pivotal role in our life~\cite{chan2020biosignal,Hau2024Towards}, particularly for people with memory impairments~\cite{roberts2023knowledge}. With LLMs demonstrating strong language understanding and mass information processing capabilities~\cite{lewis2020retrieval}, having driven innovative applications in information collection. For example, Baek et al. ~\cite{baek2024knowledge} applied LLMs to analyze users' search engine interaction history as contextual data, providing more personalized web search results. However, this work was limited to traditional desktop interactions without real-world environment integration. Cai et al.~\cite{cai2025aiget} further identified interesting but overlooked environmental information through an LLM-empowered system combining explicit and implicit initial triggers.

Information organization constitutes another crucial component in knowledge augmentation systems. Yu et al.~\cite{yu2024bnotehelper} proposed Bnotehelper, which uses LLMs to rapidly generate note outlines from video content. However, this process neglected the importance of human-in-the-loop mechanisms. Yen et al. ~\cite{yen2024memolet} tackle organizational challenges by transforming static LLM conversation histories into interactive object. 

LLMs should also assist people in naturally utilizing stored information during daily life. Zulfikar et al. 
     ~\cite{zulfikar2024memoro} made progress by implementing LLMs that anticipate users' memory needs through conversational context analysis, reducing retrieval effort. However, their framework focuses exclusively on textual information, neglecting the growing prevalence of visual data like images and videos in personal knowledge systems~\cite{hu2025vision}.

\section{Think-Aloud Study}
\subsection{Study Design}
To explore how users interact with and record information in a real-world context, we conducted a think-aloud study in an exhibition setting. Participants were tasked with exploring the exhibition, where they encountered various types of information, and were given the flexibility to define and experiment with any interaction mode they deemed suitable for capturing information. Throughout the process, we recorded their interactions via video, capturing what types of information they chose to record, the methods they employed for documentation, and their cognitive processes in interpreting the content. 

\subsubsection{Exhibition Setting}

The study was conducted in the visitor center of our institution, an information-rich exhibition space designed to showcase the university's research, academic achievements, and institutional history. The exhibition featured a diverse range of interaction channels, including virtual reality (VR), gesture-based interfaces, desktop-based interactions, museum-like installations, video and audio content, and immersive experiences. This multi-modal environment provided a complex, information-intensive setting where participants engaged with dense academic material through various sensory and interactive modalities. 

Participants were required to navigate, filter, and synthesize information across multiple formats, often switching between passive consumption (e.g., reading and watching videos) and active interaction (e.g., VR exploration and gesture-based engagement). Additionally, the setting introduced contextual challenges such as time constraints, selective attention demands, and varying levels of engagement across different media. By analyzing participants' behaviors in this environment, we aimed to uncover how individuals process and retain complex academic content and how AI-driven cognitive augmentation can enhance structured reasoning and information synthesis.

\begin{figure}
    \centering
    \includegraphics[width=0.7\linewidth]{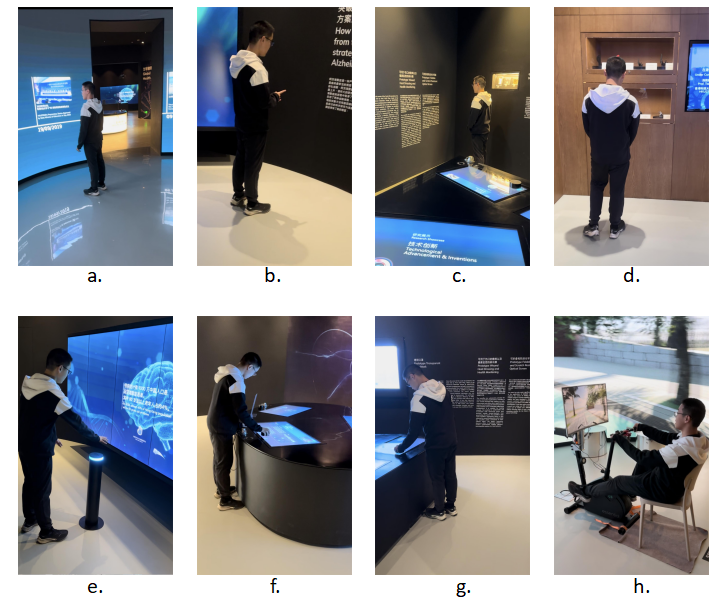}
    \caption{Overview of the exhibition setup, which includes different modalities of information presentation and interaction. The exhibition integrates various digital and physical interfaces, including large-scale digital displays for visual information (a), textual and interactive tabletop interfaces for content exploration (c, f, g), object-based museum-like exhibits (d), and sensor-based gesture interaction (e). Additionally, there are also immersive simulations (h), which provides an overview of the campus resources and map. The information is mainly composed of academic content (b, c, d, e, f, g) and school information (a, h)}
    \label{fig:enter-label}
\end{figure}

\subsubsection{Participants}
We recruited three participants, all of whom were MPhil or PhD students from the university. Each participant had at least five years of experience in design or development, making them well-versed in knowledge-intensive tasks and interactive systems. Their background ensured they were familiar with information structuring, digital interaction, and knowledge processing, making their insights valuable for understanding cognitive augmentation needs.

\subsubsection{Procedure}
Before beginning the study, participants were informed that they would need to write a report summarizing key insights from their exhibition visit. This task was designed to ensure they had a concrete goal while navigating the space, encouraging natural engagement with the information. 

The study followed a think-aloud protocol, where participants freely explored the exhibition while verbalizing their thoughts, strategies, and challenges in processing information. They were allowed to interact with various media, including text, videos, VR, and gesture-based interfaces, and to record notes in any format they preferred. Their interactions were observed and recorded for post-analysis. After the exhibition walkthrough, a semi-structured interview was conducted to capture reflections on their experience, challenges in synthesizing information, and expectations for AI-driven cognitive support. 

This study was conducted in compliance with institutional ethical guidelines and received approval from the university's Institutional Review Board (IRB). Participants provided informed consent before participation, and all data were anonymized to protect their privacy.

\subsection{Preliminary Findings}

\subsubsection{Participants' Information Processing Strategies}
Participants processed information through multiple modalities, including text, video, and interactive displays. They used textual content for comprehension and later reference, with some reading aloud to reinforce understanding and others storing text for post-visit review. They treated video recordings as a backup for later reflection rather than an immediate resource. They captured photographs to document key information but struggled with organization and retrieval. Two of the participants (N=2/3) mentioned the intention to annotate images in real-time.

Participants unconsciously collected additional information through movement patterns, navigation sequences, and habitual gestures such as finger-pointing and silent reading. For example, a participant is habitually pointing and silently reading the content they are interested to. They took personal notes that focused more on subjective impressions than on structured content, making later recall difficult. These behaviors showed that participants engaged in both explicit and implicit information processing, requiring AI support to help structure and retrieve knowledge effectively.

\subsubsection{Environmental and Social Constraints in Context-Aware AI Interaction}
Participants adjusted their engagement strategies based on environmental and social constraints. They hesitated to provide real-time verbal commentary due to privacy concerns. They avoided taking photographs in dimly lit areas to prevent disturbing others. They minimized interactions requiring overt gestures or spoken input to maintain discretion in a shared public setting.

These behaviors demonstrated a major limitation in existing AI-driven cognitive augmentation systems. AI systems fail to recognize social and environmental contexts, leading to rigid or intrusive interactions. Participants needed AI support that could adapt to their surroundings, offering subtle interaction methods such as silent note-taking suggestions, discreet notifications for summarization, and adaptive interfaces that adjusted based on the environment. The exhibition setting also introduced cognitive load challenges, requiring participants to selectively focus on key information while filtering distractions. AI tools need to track users' cognitive states and provide timely, relevant support, such as summarizing dense content or reinforcing key insights.

\subsubsection{Participants' Expectations for Adaptive AI Support}
Participants identified two key areas where AI could improve their experience: (1) real-time comprehension support and (2) post-visit knowledge organization. They needed AI systems to help process and contextualize information as they explored, rather than requiring them to manually curate their knowledge afterward. They also expected AI to assist in aggregating, organizing, and structuring recorded content for later reference.

Participants demonstrated different cognitive workflows for structuring information. A participant constructed broad conceptual frameworks before revisiting specific details. Two others adopted a detail-oriented approach, capturing comprehensive information first before synthesizing key insights. These differences showed that rigid, one-size-fits-all AI models failed to support diverse cognitive needs. AI-driven augmentation need to recognize user-specific cognitive patterns and dynamically adjust interventions—whether by prompting reflective questions, restructuring information hierarchically, or offering personalized synthesis tools.

\subsection{Implications for Context-Aware Cognitive Augmentation}
Participants' behaviors suggest the potential benefits of \textbf{multi-modal, context-sensitive, and adaptive AI systems} in supporting human cognition. Existing AI models primarily rely on reactive interactions, requiring users to explicitly request information~\cite{zhang2024debates}. However, our findings indicate that AI systems could be more effective if they \textbf{adapted} to users' cognitive states and environments to provide more relevant support.

Based on these observations, we outline key considerations for designing AI-driven cognitive augmentation systems:
\begin{itemize}
    \item \textbf{Multi-modal Awareness}: AI could integrate text, images, movement patterns, and behavioral cues to enhance contextualized knowledge augmentation.
    \item \textbf{Cognitive Workflow Adaptation}: AI might benefit from recognizing whether a user is engaging in exploratory or structured information processing and adjusting its interventions accordingly.
    \item \textbf{Socially Adaptive Interaction}: AI should consider offering discreet, unobtrusive engagement methods that respect social constraints while assisting cognition.
    \item \textbf{Seamless Transition Between Real-Time and Long-Term Support}: AI could function both as a real-time cognitive assistant and a post-experience knowledge synthesizer.
\end{itemize}

\section{Conclusion}
Our study highlights the critical need for AI systems that go beyond static, prompt-driven interactions to proactively assist users in sensing, processing, and utilizing information in real-world tasks by adapting to cognitive constraints and integrating seamlessly into daily workflows. By examining how individuals engage with information in real-world settings, we identified key challenges in context-aware knowledge capture, information organization, and adaptive cognitive support. Our findings reinforce that effective AI-driven cognitive augmentation must be multi-modal, personalized, and socially adaptive—capable of understanding user behaviors, guiding engagement dynamically, and transitioning seamlessly between real-time assistance and long-term cognitive scaffolding. Building upon these insights, future research should focus on developing structured user-context knowledge spaces that empower LLMs to function as proactive collaborators rather than passive responders. By incorporating context-awareness, proactive engagement, and structured knowledge integration, AI systems can move closer to fostering meaningful, personalized cognitive augmentation.

\begin{acks}
We thank CCF-Lenovo Research Fund (Grant No. 20240102) for supporting this work. We also acknowledge the support from Guangdong Provincial Key Lab of Integrated Communication, Sensing and Computation for Ubiquitous Internet of Things (No.2023B1212010007).
\end{acks}

\bibliographystyle{ACM-Reference-Format}
\bibliography{sample-base}

\appendix
\section{Semi-Structured Interview Guide}

This appendix presents the semi-structured interview guide used in our study. 

\subsection{Information Needs}
\begin{itemize}
    \item What specific information is essential for writing about the exhibition?
    \item How do you determine what data to collect during the visit?
\end{itemize}

\subsection{Interaction Preferences}
\begin{itemize}
    \item What is the best way to interact with the data collection system?
    \item Are certain interactions more intuitive or effective?
    \item What are the public concerns regarding interaction with such a system?
\end{itemize}

\subsection{Data Organization}
\begin{itemize}
    \item How should the system organize the collected data?
    \item Should there be a data organization/reorganization step? How do you prefer to participate in this step?
    \item Do you prefer manual involvement in organizing the data, or should AI handle this step?
\end{itemize}

\subsection{Collaboration with AI}
\begin{itemize}
    \item What role should AI play in assisting with organizing data during writing/preparing for presentation?
    \item How can collaboration between the user and AI be improved?
\end{itemize}

\subsection{Information Triggers}
\begin{itemize}
    \item What types of information should be captured during the tour?
    \item Are there specific triggers or guidelines for initiating data collection?
    \item Do you think there are some guidelines regarding the information collection/trigger process?
\end{itemize}

\end{document}